\documentclass[useAMS,usegraphicx,usenatbib]{mn2e}

\newcommand{\kms}{\,km\,s$^{-1}$}                          
\newcommand{\ion}[2]{{#1}\,{\sc {\small{#2}}}}             
\newcommand{\Porb}{\ensuremath{P_{\rm orb}}}               
\newcommand{\Pspin}{\ensuremath{P_{\rm spin}}}             
\newcommand{\mc}[1]{\multicolumn{2}{c}{#1}}

\title[The intermediate polar SDSS\,J233325.92$+$152222.1]
      {SDSS\,J233325.92$+$152222.1 and the evolution of intermediate polars}

\author[Southworth et al.]
       {John Southworth$^1$
        \thanks{E-mail: j.k.taylor@warwick.ac.uk (JS),
        \newline Boris.Gaensicke@warwick.ac.uk (BTG),
        \newline T.R.Marsh@warwick.ac.uk (TRM)},
        B.\ T.\ G\"ansicke$^1$,
        T.\ R.\ Marsh$^1$,
        D.\ de Martino$^2$, \newauthor
        A.\ Aungwerojwit$^{1,3}$ \\
        $^1$ Department of Physics, University of Warwick, Coventry, CV4 7AL, UK          \\
        $^2$ INAF -- Osservatorio di Capodimonte, Via Moiariello 16, 80131 Napoli, Italy  \\
        $^3$ Department of Physics, Faculty of Science, Naresuan University, Phitsanulok, 60500, Thailand}

\begin{document} \maketitle 

\begin{abstract}
Intermediate polars (IPs) are cataclysmic variables which contain magnetic white dwarfs with a rotational period shorter than the binary orbital period. Evolutionary theory predicts that IPs with long orbital periods evolve through the 2--3\,hr period gap, but it is very uncertain what the properties of the resulting objects are. Whilst a relatively large number of long-period IPs are known, very few of these have short orbital periods. We present phase-resolved spectroscopy and photometry of SDSS\,J233325.92$+$152222.1 and classify it as the IP with the shortest known orbital period ($83.12 \pm 0.09$\,min), which contains a white dwarf with a relatively long spin period ($41.66 \pm 0.13$\,min). We estimate the white dwarf's magnetic moment to be $\mu_{\rm WD} \approx 2 \times 10^{33}$\,G\,cm$^{3}$, which is not only similar to three of the other four confirmed short-period IPs but also to those of many of the long-period IPs. We suggest that long-period IPs conserve their magnetic moment as they evolve towards shorter orbital periods. Therefore the dominant population of long-period IPs, which have white dwarf spin periods roughly ten times shorter than their orbital periods, will likely end up as short-period IPs like SDSS\,J2333, with spin periods a large fraction of their orbital periods.
\end{abstract}


\section{Introduction}                                                                       \label{sec:intro}

Cataclysmic variables (CVs) are interacting binary stars containing a white dwarf primary star and a low-mass secondary star in a tight orbit \citep{Warner95book}. In most of these systems the secondary star is unevolved and fills its Roche lobe, losing material to the white dwarf primary via an accretion disc. About a quarter of CVs contain magnetic white dwarfs, and these systems are split into two categories: the AM\,Her stars (polars) and the intermediate polars (IPs). In polars, the strong magnetic field of the white dwarf causes its spin period, \Pspin, to synchronise to the orbital period, \Porb. The evolution of these objects is expected to be strongly affected by interaction between the magnetic fields of the white dwarf and the late-type secondary star \citep{WebbinkWick02mn}. In IPs it is believed that the lower magnetic field of the white dwarf means that its rotation has not become synchronised to the orbital motion. The evolution of IPs and in particular their relation to polars is still a debated issue (\citealt{Hellier01book}; \citealt{Norton++04apj}, hereafter NWS04).

While the total number of confirmed IPs has been boosted substantially over the past few years \citep[e.g.][]{WoudtWarner03mn, Araujo+03aa, Rodriguez+04mn, Woudt++04mn, Schlegel05aa, Rodriguez+05aa, Gansicke+05mn, Bonnet+06aa, NortonTanner06aa, Demartino+06aa}, only a handful of these have short orbital periods (i.e., below the 2--3\,hr gap which is prevalent in the observed population of CVs). Conversely, the known population of polars mostly lies below $\Porb = 4$\,hr \citep{WebbinkWick02mn}. This has led to suggestions that long-period IPs may evolve into short-period polars \citep[see][]{Patterson94pasp}. A detailed theoretical study by NWS04 found that in general this should be the case. However, those with a white dwarf magnetic moment of $\mu_{\rm WD} \la 5 \times 10^{33}$\,G\,cm$^3$ {\em and} secondary stars with weak magnetic fields will remain IPs, because the magnetic interaction between the stars is not strong enough to synchronise the the rotation of the white dwarf to the orbital motion. This picture of the evolution of magnetic CVs has not yet been confirmed observationally because very few short-period IPs are known.

\begin{figure*} \includegraphics[width=\textwidth,angle=0]{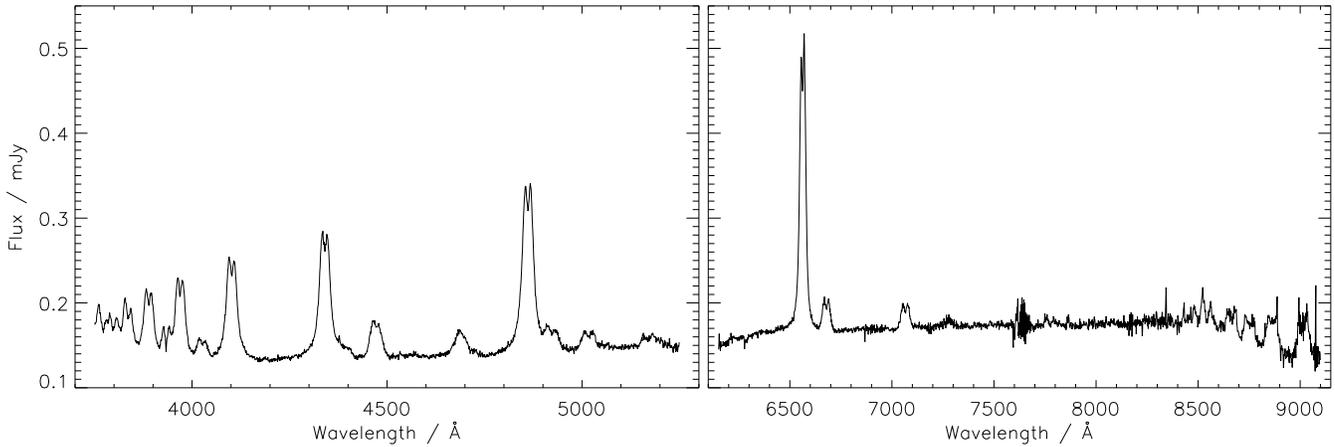}
\caption{\label{fig:whtspec} Flux-calibrated averaged blue-arm (left) and
red-arm (right) WHT spectra of SDSS\,J2333.} \end{figure*}

We are conducting a research program \citep{Gansicke05aspc} to measure the orbital periods of objects spectroscopically identified as CVs \citep{Szkody+02aj, Szkody+03aj, Szkody+04aj, Szkody+05aj, Szkody+06aj}
by the Sloan Digital Sky Survey (SDSS). The main motivation of this work lies in the characterisation of an homogeneously identified sample of CVs which can be used to investigate the properties of the intrinsic CV population. The relatively faint limiting magnitude of the SDSS spectroscopic observations means that this survey is much less biased towards intrinsically brighter objects than previous work. A very high proportion of the SDSS CVs which we have studied so far are faint short-period systems (\citealt{Gansicke+06mn}; \citealt[][hereafter Paper\,I]{Me+06mn}; \citealt{Littlefair+06sci}). In Paper\,I we found that SDSS\,J023322.61+005059.5 was a probable IP with $\Porb = 96.08 \pm 0.09$\,min and $\Pspin \approx 60$\,min. Here we present a detailed analysis of another system, SDSS\,J233325.92$+$152222.1 (hereafter SDSS\,J2333), which we find to be an IP with $\Porb = 83.12$\,min and $\Pspin = 41.66$\,min\footnote{The reduced spectra and photometry presented in this work will be made available at the CDS ({\tt http://cdsweb.u-strasbg.fr/}) and at {\tt http://www.astro.keele.ac.uk/$\sim$jkt/}}.

\citet{Szkody+05aj} presented the identification spectrum of SDSS\,J2333 alongside time-resolved photometry and spectroscopy. They found modulation at 21\,min in the light curve and at 82\,min in the H$\alpha$ and H$\beta$ emission line radial velocities, and suggested that this object could be an IP with an orbital period close to the 80\,min period minimum for CVs \citep{Knigge06mn}. The SDSS spectrum shows the Balmer and \ion{He}{I} emission lines characteristic of CVs as well emission at \ion{He}{II} 4686\,\AA\ which is a hallmark of magnetic systems. Whilst the Balmer and \ion{He}{I} lines have the double-peaked profile indicative of an accretion disc, the \ion{He}{II} line is single-peaked. However, it is difficult to conclude anything from this because the line may be contaminated by emission from the \ion{C}{III} and \ion{N}{III} Bowen blend. X-ray emission from SDSS\,J2333 was not detected by ROSAT.


\begin{table*} \begin{center}
\caption{\label{tab:obslog} Log of the observations of SDSS\,J2333.}
\begin{tabular}{cccccrrr} \hline
Start date & Start time & End time & Telescope and & Optical  &  Number of   & Exposure \\
(UT)       &  (UT)      &  (UT)    &  instrument   & element  & observations & time (s) \\
\hline
2006 09 13 & 19 57 & 23 58 & CA 2.2 CAFOS & (unfiltered)    & 293 &  40 \\
2006 09 25 & 00 46 & 04 46 & WHT ISIS     & R600B R316R     &  22 & 600 \\
2006 09 25 & 23 07 & 01 39 & WHT ISIS     & R600B R316R     &  15 & 600 \\
\hline \end{tabular} \end{center} \end{table*}


\begin{figure}
\includegraphics[width=0.23\textwidth,angle=0]{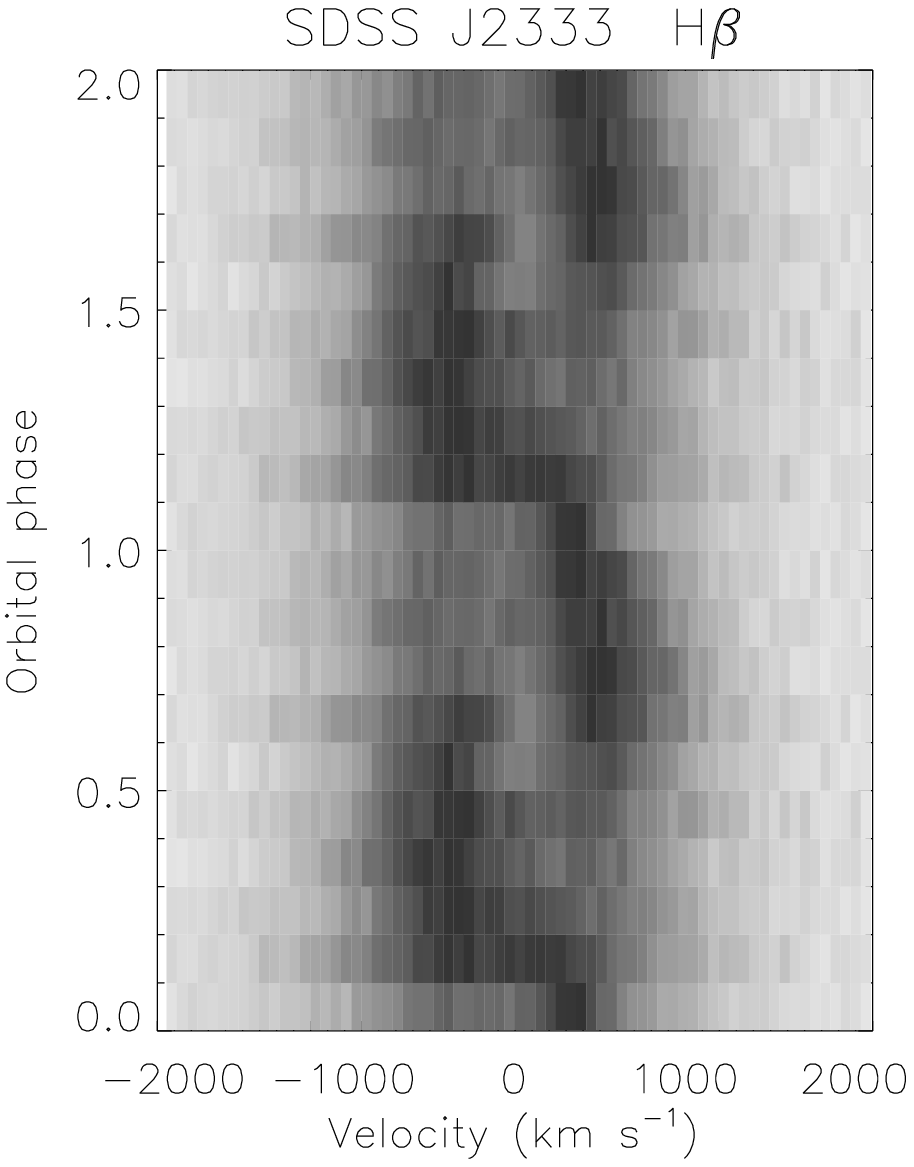}
\includegraphics[width=0.23\textwidth,angle=0]{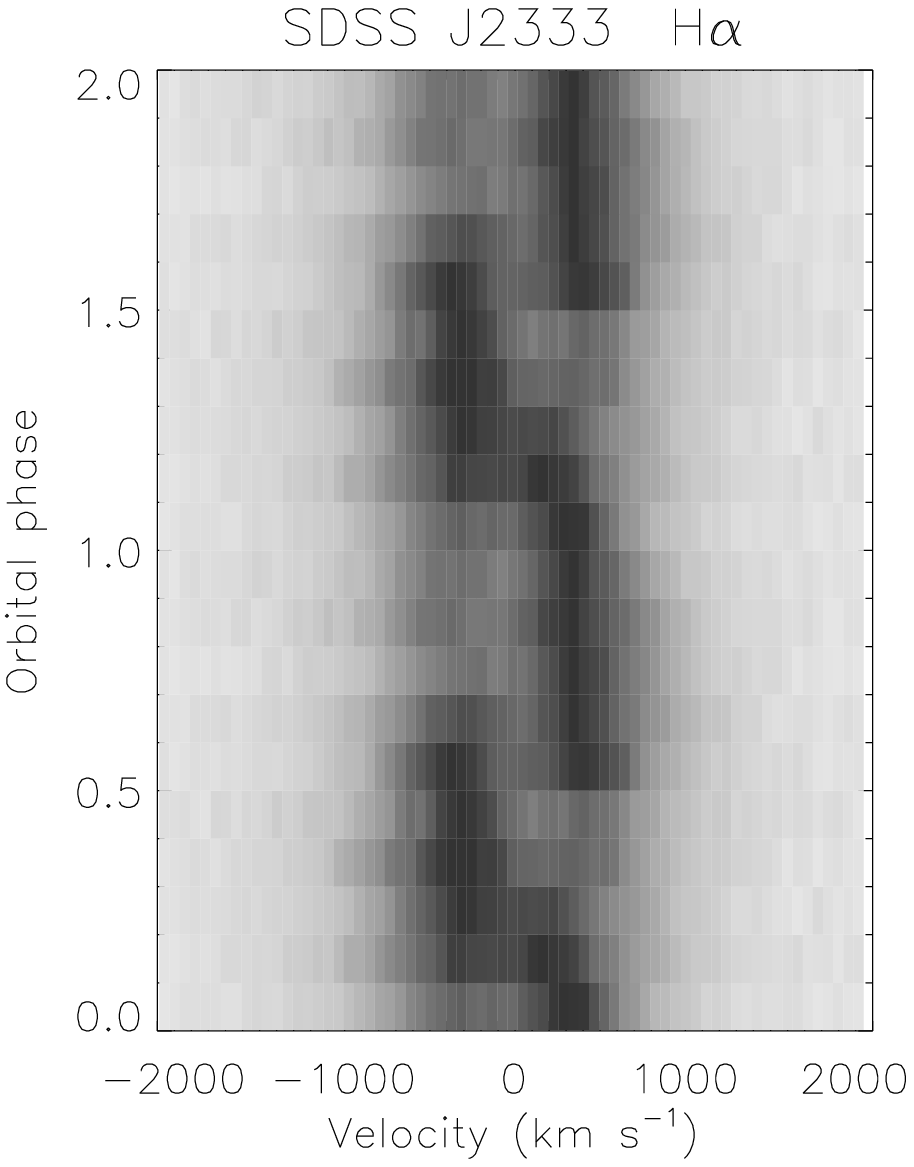}
\caption{\label{fig:trailed} Greyscale plots of the continuum-normalised
and phase-binned trailed spectra around the H$\alpha$ and H$\beta$
emission lines. Other Balmer lines give similar result, but trails for
the helium emission lines are too noisy to be useful.}\end{figure}

\section{Observations and data reduction}\label{sec:obs}

\subsection{WHT spectroscopy}                                                       \label{sec:obs:whtspec}

\begin{table*} \begin{center}
\caption{\label{tab:orbits} Results of the bootstrapping analysis and the best-fitting
circular spectroscopic orbits for the H$\alpha$ line, both radial velocity measurement
techniques and for different possible periods. The orbit adopted as a final result is
indicated in bold. The results of the bootstrapping analysis are given as the percentage
of simulations returning a highest peak close to a particular period.}
\begin{tabular}{l r@{\,$\pm$\,}l r@{\,$\pm$\,}l r@{\,$\pm$\,}l c r}\hline
Radial velocity &    \mc{Orbital period}   &  \mc{Velocity amplitude} &  \mc{Systemic velocity}  & $\sigma_{\rm rms}$ & Scargle bootstrap  \\
measurement     &         \mc{(day)}       &    \mc{(\kms)}     &     \mc{(\kms)}    &      (\kms)        & probability (\%)  \\
\hline
Double Gaussian &    0.051038 &    0.00011 &     51.6 &     7.8 &   $-$2.7 &     5.6 &    34 &      \\
Double Gaussian &    0.054111 &    0.00012 &     56.3 &     6.4 &   $-$0.8 &     4.6 &    28 & 11.5 \\
Double Gaussian &    0.057622 &    0.00013 &     58.4 &     6.0 &      2.6 &     4.3 &    26 & 85.6 \\
Double Gaussian &    0.061608 &    0.00016 &     57.1 &     7.0 &      6.6 &     4.9 &    30 &  2.9 \\[3pt]
Single Gaussian &    0.051149 &    0.00012 &    151.8 &    17.6 &     21.3 &    12.9 &    78 &      \\
Single Gaussian &    0.054222 &    0.00008 &    168.2 &    11.6 &     27.5 &     8.5 &    52 & 17.4 \\
Single Gaussian &{\bf0.057724}&{\bf0.00006}&{\bf176.3}&{\bf 8.8}&{\bf 38.9}&{\bf 6.4}&{\bf39}& 82.6 \\
Single Gaussian &    0.061713 &    0.00009 &    175.5 &    12.9 &     52.5 &     9.1 &    55 &      \\
\hline \end{tabular} \end{center} \end{table*}

Spectroscopic observations were obtained in 2006 September using the ISIS double-beam spectrograph on the William Herschel Telescope (WHT) at La Palma (Table\,\ref{tab:obslog}). For the red arm we used the R316R grating and Marconi CCD binned by factors of 2 (spectral) and 4 (spatial), giving a wavelength range of 6300--9200\,\AA\ and a reciprocal dispersion of 1.66\,\AA\ per binned pixel. For the blue arm we used the R600B grating and EEV12 CCD with the same binning factors, giving a wavelength coverage of 3640--5270\,\AA\ at 0.88\,\AA\ per binned pixel. From measurements of the full widths at half maximum (FWHMs) of arc-lamp and night-sky spectral emission lines, we find that this gave resolutions of 3.5\,\AA\ (red arm) and 1.8\,\AA\ (blue arm).

Data reduction was undertaken using optimal extraction (\citealt{Horne86pasp}) as implemented in the {\sc pamela}\footnote{{\sc pamela} and {\sc molly} were written by TRM and can be found at {\tt http://www.warwick.ac.uk/go/trmarsh}} code \citep{Marsh89pasp}. The wavelength calibration was interpolated from copper-neon and copper-argon arc lamp exposures taken every hour. We removed the telluric lines and flux-calibrated the target spectra using observations of G191-B2B, treating each night separately.

A total of 38 spectra were obtained over two nights and the average spectrum is shown in Fig.\,\ref{fig:whtspec}. They show strong double-peaked Balmer emission and double-peaked emission from the other usual suspects, \ion{He}{I}, \ion{Fe}{II} and \ion{Ca}{II}\,K. There is no hint of the Balmer lines of the underlying white dwarf or of \ion{Ca}{II} emission. However, there is significant {\em single}-peaked emission at \ion{He}{II} 4686\,\AA, which is characteristic of systems containing magnetic white dwarfs (e.g., DW\,Cnc, \citealt{Rodriguez+04mn}). The trailed spectra (Fig.\,\ref{fig:trailed}) show that the emission lines have double-peaked emission with a strong S-wave component. The spectrum of SDSS\,J2333 is reminiscent of those of the short-period IPs HT\,Cam \citep{Kemp+02pasp}, DW\,Cnc \citep{Rodriguez+04mn} and V1025\,Cen \citep{Buckley+98mn}. The equivalent widths of H$\alpha$, H$\beta$ and \ion{He}{II} 4686\,\AA\ are $78 \pm 4$, $56 \pm 3$ and $8 \pm 1$\,\AA, respectively. The ratio $\frac{EW({\rm 4686})}{EW({\rm H\beta})} = 0.14 \pm 0.02$ is typical for short-period IPs (e.g., DW\,Cnc) and not unusual for short-period non-magnetic CVs \citep[e.g.][]{ThorstensenFenton03pasp}.



\subsection{Calar Alto photometry}                                                        \label{sec:obs:phot}

Unfiltered differential photometry of SDSS\,J2333 was obtained on the night of 2006 September 13 using the 2.2m telescope at Calar Alto Observatory equipped with the CAFOS focal reducer (Table\,\ref{tab:obslog}). We used the blue-optimised SITe CCD without binning, giving a plate scale of 0.53''\,px$^{-1}$. The images were reduced and aperture photometry was performed using the pipeline described by \citet{gansicke+04aa}, which applies bias and flat-field corrections within {\sc midas}\footnote{\tt http://www.eso.org/projects/esomidas/} and uses the {\sc sextractor} package \citep{BertinArnouts96aas} to obtain aperture photometry for all objects in the field of view. The differential magnitudes were converted into apparent magnitudes using the SDSS $g$ and $r$ photometry of the main comparison star.


\section{Data analysis}

\subsection{Radial velocity analysis}

We measured radial velocities from the Balmer emission lines by cross-correlation with single and double Gaussian functions \citep{SchneiderYoung80apj}. For each line we tried a range of different widths and separations for the Gaussians in order to verify the consistency of our results (see Paper\,I for further details). We found that the best results for period determination were obtained using the H$\alpha$ line and a single Gaussian of FWHM 1000\kms, as expected given the prominence of the S-wave emission in the spectra (Fig.\,\ref{fig:trailed}). For a double Gaussian the best parameters were FWHM 300\kms\ and separation 2000\kms. The choice of emission line, measurement technique and Gaussian parameters does not have a significant effect on the derived period. Radial velocities derived for the \ion{He}{I} 4471\,\AA\ line gave similar results but with much larger noise as the line is weaker.

\begin{figure} \includegraphics[width=0.48\textwidth,angle=0]{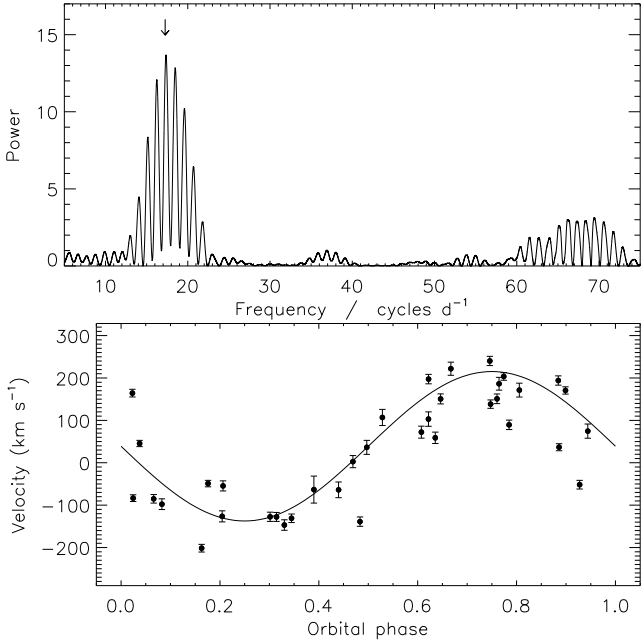} \\
\caption{\label{fig:rvorbit} Scargle periodogram of the H$\alpha$ radial velocities
measured using a single Gaussian of width 1000\kms\ (upper panel) and the phased
velocities with the best-fitting circular spectroscopic orbit (lower panel). On the
periodogram the chosen peak is indicated with an arrow.} \end{figure}

Periodograms were calculated from the measured H$\alpha$ radial velocities using the \citet{Scargle82apj} method, analysis of variance \citep[AoV; ][]{Schwarzenberg89mn} and orthogonal polynomials \citep[ORT; ][]{Schwarzenberg96apj}, as implemented within the {\sc tsa}\footnote{\scriptsize\tt http://www.eso.org/projects/esomidas/doc/user/98NOV/volb/node220.html} context in {\sc midas}. All three types of periodogram contain a strong signal close to  83\,min, accompanied by one-day aliases, and a few small peaks at much shorter periods (Fig.\,\ref{fig:rvorbit}). As the three periodogram techniques agree well we have restricted further analysis to the Scargle algorithm, which is the most appropriate for a simple sinusoidal variation. We obtained the orbital period by fitting a circular spectroscopic orbit (sine wave) to the radial velocities (Fig.\,\ref{fig:rvorbit}) using the {\sc sbop}\footnote{Spectroscopic Binary Orbit Program, written by P.\ B.\ Etzel, \\ {\tt http://mintaka.sdsu.edu/faculty/etzel/}} program, which we have previously found to give reliable error estimates for the optimised parameters \citep{Me+05mn}. Whilst the radial velocities measured using a single Gaussian have a larger scatter than those from the double-Gaussian technique, the velocity variation in the core of the emission lines is much larger. These radial velocities therefore give the best results, and we find a period of $\Porb = 83.12 \pm 0.09$\,min.

To investigate the probability that our chosen alias is the correct one we have both fitted sine curves for each of the possible alias periods and conducted bootstrapping simulations (see Paper\,I). For the sine curves we would expect the lowest errors and residuals, and the highest amplitude, for the correct period. The bootstrapping gives results which can be directly interpreted as a probability distribution, but there will be some bias away from the correct period because the process of resampling the observations leads to a loss of temporal resolution, degrading the resulting periodograms. Thus bootstrapping will underestimate the probability that the best period is in fact the correct one. Both the sine curve fitting and the bootstrapping simulations clearly favour the 83.12\,min period.

\subsection{Light curve analysis}

\begin{figure} \includegraphics[width=0.48\textwidth,angle=0]{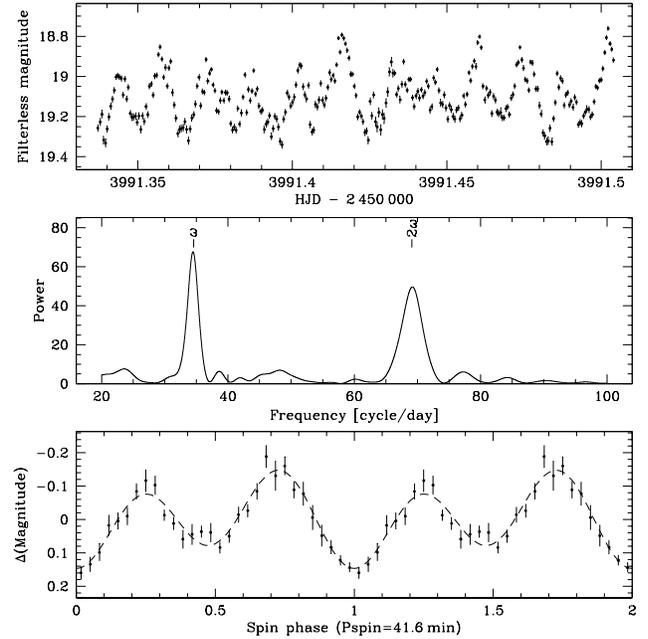}\\
\caption{\label{fig:2333:lcplot} Calar Alto 2.2m photometry of SDSS\,J2333. The top
panel shows the observed light curve, which displays a clear sinusoidal variation.
The middle panel contains an ORT periodogram and the lower panel depicts the phased
light curve with a double-sine fit.} \end{figure}

The light curve of SDSS\,J2333 shows clear sinusoidal variation with a large amplitude of about 0.2\,mag (Fig.\,\ref{fig:2333:lcplot}). An ORT periodogram of these data has strong power at one half and one quarter of the orbital period (Fig.\,\ref{fig:2333:lcplot}) but not at the orbital period itself. A visual inspection of the light curve suggested that it exhibits a double-humped variation so we fitted the function
\[ f(x) = a_1 \sin \frac{2\pi(x+\phi_1)}{P} + a_2 \sin \frac{2\pi(2x+\phi_2)}{P} \]
where the optimised parameters are $a_1$ and $a_2$ (the amplitudes of variation), $P$ (the period) and $\phi_1$ and $\phi_2$ (the phase zeropoints). This process confirmed the double-humped nature of the variation (Fig.\,\ref{fig:2333:lcplot}) and yielded a period of $41.66 \pm 0.13$\,min.

\citet{Szkody+05aj} found a periodicity of 21\,min in their light curve of this system, which was obtained with a much smaller telescope (1.0\,m) than our data. Aside from the fact that the double-peaked nature of the variation was not apparent in their data, this periodicity is consistent with our own measurement and demonstrates that it is stable.

We have found that SDSS\,J2333 has $\Porb = 83.12 \pm 0.09$\,min, close to the minimum orbital period for CVs, and a light curve with a strong double-humped modulation of amplitude 0.2\,mag and period $41.66 \pm 0.13$\,min. We interpret the light variation to be due to accretion heated spots on the white dwarf primary, which has a largely bipolar magnetic field. Therefore we classify SDSS\,J2333 as an IP with an orbital period below the period gap and white dwarf which rotates with twice the frequency of the orbital motion.


\section{Discussion}                                                                    \label{sec:discussion}

\begin{table*} \begin{center}
\caption{\label{tab:IPperiod} Orbital and spin periods of intermediate polar CVs
with orbital periods shorter than the 2--3\,hr period gap. The estimated magnetic
moments of the systems are from NWS04, either taken from their Table\,1 or traced
from their Fig.\,2.
\newline $^*$ The IP nature of RX\,J1039$-$0507 and SDSS\,J023322.61+005059.5 has not been confirmed.
\newline {\bf References:}  (1) \citet{Tovmassian+98aa}; (2) \citet{Kemp+02pasp};
(3) \citet{Demartino+05aa}; (4) \citet{Rodriguez+04mn};  (5) \citet{Patterson+04pasp};
(6) \citet{Buckley+98mn};   (7) \citet{Hellier++98mn};   (8) \citet{Hellier++02mn};
(9) \citet{Mumford67apjs}; (10) \citet{Vogt++80aa};     (11) \citet{HellierSproats92ivbs};
(12) \citet{WoudtWarner03mn}.}
\begin{tabular}{l r@{\,$\pm$\,}l r@{\,$\pm$\,}l c c l} \hline
System               &\mc{Orbital period (min)}& \mc{Spin period} & $\underline{\Pspin}$ &  Estimated magnetic  & References\\
                     &       \mc{(min)}        &    \mc{(min)}    & \Porb  &  moment (G\,cm$^3$)  &           \\
\hline
HT Cam               &  85.9853   & 0.0014     & 8.58430 &0.00000 & 0.100  & 0.3$\times$10$^{33}$ & 1, 2, 3   \\
DW Cnc               &  86.1015   & 0.0003     & 38.58377&0.00006 & 0.448  & 1.5$\times$10$^{33}$ & 4, 5      \\
V1025 Cen            &      \mc{84.62}         & 35.73   & 0.05   & 0.422  & 1.8$\times$10$^{33}$ & 6, 7, 8   \\
EX Hya               &  98.256738 & 0.000006   & 67.02688&0.00001 & 0.682&$\sim$5$\times$10$^{33}$& 9, 10, 11 \\
SDSS\,J2333          &  83.12     & 0.09       & 41.66   & 0.13   & 0.501&$\sim$2$\times$10$^{33}$& This work \\
RX\,J1039.7$-$0507 $^*$&94.4597   & 0.0001     & \mc{24.062 $^*$} & 0.255  & 0.9$\times$10$^{33}$ & 12        \\
SDSS\,J023322.61+005059.5 $^*$     &  96.08     & 0.09       & 60      & 5 $^*$ & 0.625  &                      & Paper\,I  \\
\hline \end{tabular} \end{center}
\end{table*}

\begin{figure} \includegraphics[width=0.48\textwidth,angle=0]{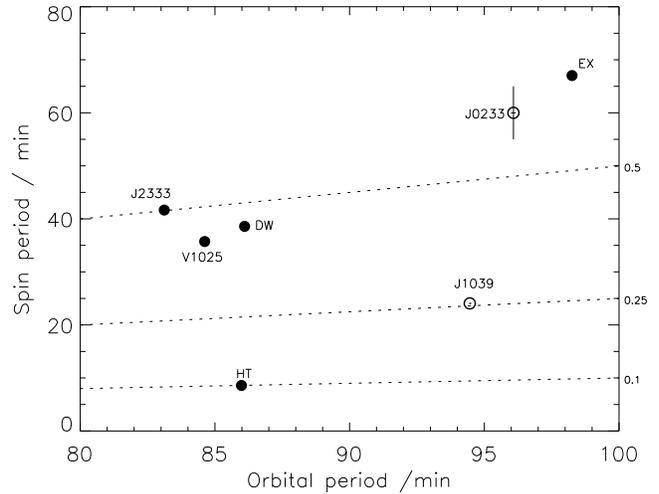} \\
\caption{\label{fig:2333:IPplot} Comparison of the orbital and spin periods
of IPs with short orbital periods. Dotted lines indicate where the spin period is
0.1, 0.25 and 0.5 times the orbital period. Those systems which are not confirmed IPs
are plotted with open circles. The error bars are smaller than the points
for all systems except SDSS\,J023322.61+005059.5.} \end{figure}

In Table\,\ref{tab:IPperiod} we have collected the observed orbital and spin periods for the five confirmed IPs with $\Porb < 2$\,hr. These are plotted in Fig.\,\ref{fig:2333:IPplot}. Classification as an IP requires the presence of coherent variation at the white dwarf spin period over a significant span of time \citep[e.g.][]{Buckley00newar}. As the photometric observations for an additional two objects, RX\,J1039.7$-$0507 \citep{WoudtWarner03mn} and SDSS\,J023322.61+005059.5 (Paper\,I), have only a short baseline their IP nature needs additional confirmation. The white dwarf magnetic moments, $\mu_{\rm WD}$, of the objects in Table\,\ref{tab:IPperiod} have been obtained from NWS04 under the assumption that the systems are in rotational equilibrium, either taken from their Table\,1 or estimated from their Fig.\,2.

SDSS\,J2333 is joined by the objects DW\,Cancri and V1025\,Centauri in having $\Pspin \approx 0.5 \Porb$ and $\mu_{\rm WD} \sim 2 \times 10^{33}$\,G\,cm$^3$. Assuming a canonical white dwarf radius of 10$^7$\,m, this gives a field strength of 2\,MG for the white dwarf. The properties of EX\,Hya are also similar, whereas those of the fifth system, HT\,Cam, are quite different.

The majority of IPs with $\Porb > 3$\,hr have $\Pspin \approx 0.1 \Porb$ \citep{Barrett++88mn,Gansicke+05mn} and $\mu_{\rm WD} \sim 2 \times 10^{33}$\,G\,cm$^3$ (NWS04). This magnetic moment is strikingly similar to those of SDSS\,J2333, DW\,Cnc and V1025\,Cen. This is an empirical indication that long-period IPs with $\Pspin \approx 0.1 \Porb$ conserve $\mu_{\rm WD}$ as they evolve and become short-period IPs with $\Pspin \approx 0.5 \Porb$. This is consistent with the evolutionary picture of magnetic CVs drawn by \citet{Patterson94pasp,WebbinkWick02mn} and NWS04.

HT\,Cam is a short-period IP with a lower $\mu_{\rm WD}$ of $2.7 \times 10^{32}$\,G\,cm$^3$. In the scenario we have outlined, an object such as this will result from the evolution of systems which currently resemble SDSS\,J223843.84$+$010820.7 (Aqr\,1 in \citealt{Woudt++04mn}) for which $\mu_{\rm WD} = 2.6 \times 10^{32}$\,G\,cm$^3$ has been estimated (NWS04). A similar argument can be advanced for RX\,J1039.7$-$0507, which has $\mu_{\rm WD} \approx 10^{33}$ which is very similar to that of several known IPs (NWS04), for example 1RXS\,J062518.2+733433 \citep{Staude+03aa}.

\citet{KingWynn99mn} have found that there is a large continuum of equilibrium spin levels for short-period IPs, and that an IP with a specific mass ratio, mass transfer rate and orbital period can have any of a wide range of values of \Pspin/\Porb, depending on the value of $\mu_{\rm WD}$

The theoretical calculations of NWS04 suggest that long-period IPs will evolve into polars unless $\mu_{\rm WD} \la 5 \times 10^{33}$\,G\,cm$^3$ and the magnetic field of the secondary star is weak. In this case they may become EX\,Hya-like systems as the interaction between the magnetic fields is too weak to synchronise the rotation of the white dwarf to the orbital motion.

Whilst there are twelve long-period IPs listed in NWS04 with $\mu_{\rm WD}$ similar to that of SDSS\,J2333, there are only four short-period systems with this property. As short-period CVs are predicted to be intrinsically far more common than long-period ones \citep{Dekool92aa,Kolb93aa,Politano96apj}, this means either that the vast majority of the long-period IPs become polars or that the known population of short-period IPs is much less complete than for those with longer periods. The latter possibility could easily arise as short-period IPs are in general intrinsically fainter than long-period IPs \citep{Warner95book}.


\section{Conclusions}                                                                   \label{sec:conclusion}

SDSS\,J2333 was identified as a cataclysmic variable from a spectrum taken by the SDSS, which shows strong double-peaked Balmer emission, double-peaked \ion{He}{I} emission and the single-peaked \ion{He}{II} emission which is often found in CVs containing a magnetic white dwarf. We have measured its orbital period to be $83.12 \pm 0.09$\,min from a radial velocity analysis of the Balmer emission lines. It is therefore another SDSS-identified CV with an orbital period close to the minimum period for CVs containing unevolved mass donor stars. Its light curve shows a strong variability with a double-humped nature and a period of $41.66 \pm 0.13$\,min, which is precisely half that of its orbital period. We interpret this as arising from hot spots on the surface of the white dwarf primary component caused by accretion of matter controlled by a mostly dipolar magnetic field. SDSS\,J2333 is therefore the shortest-period example of a relatively rare class of short-period intermediate polars. An X-ray detection of its spin period is highly desirable to further investigate its IP nature. The fact that its spin period is precisely half of its orbital period suggests the presence of a physical mechanism which is maintaining this as an equilibrium state.

Four out of a total of five confirmed short-period IPs, including SDSS\,J2333, have a spin period of approximately half their orbital period. These systems contain a white dwarf with a magnetic moment of $\mu_{\rm WD} \sim 2 \times 10^{33}$\,G\,cm$^3$ (corresponding to a field strength of about 2\,MG). A relatively large number of long-period ($\Porb \ga 3$\,hr) IPs have spin periods close to a tenth of their orbital periods and $\mu_{\rm WD} \sim 2 \times 10^{33}$\,G\,cm$^3$. From this we suggest that the population of long-period IPs with $\Pspin \sim 0.1 \Porb$ will conserve $\mu_{\rm WD}$ during their later evolution and become short-period IPs with $\Pspin \sim 0.5 \Porb$.


\section*{Acknowledgements}

This work is based on observations made with the William Herschel Telescope, which is operated on the island of La Palma by the Isaac Newton Group in the Spanish Observatorio del Roque de los Muchachos (ORM) of the Instituto de Astrof\'{\i}sica de Canarias (IAC).

JS acknowledges financial support from PPARC in the form of a postdoctoral research assistant position. BTG acknowledges financial support from PPARC in the form of an advanced fellowship. TRM was supported by a PPARC senior fellowship during the course of this work.
DDM acknowledges financial support from the Italian Ministry of University and Research (MIUR).
AA thanks the Royal Thai Government for a studentship.
We thank the referee for a positive report.





\label{lastpage}

\end{document}